\documentclass[twocolumn]{emulateapj}

\usepackage{graphicx}
\usepackage{latexsym}
\usepackage{amsfonts,amsmath,amssymb}
\usepackage{url}
\usepackage{xspace}
\usepackage[utf8]{inputenc}
\usepackage{fancyref}
\usepackage{hyperref}
\usepackage{etoolbox}
\hypersetup{colorlinks=false,pdfborder={0 0 0},}

\shorttitle{}
\shortauthors{Johnson et al.}
\slugcomment{Accepted to ApJ, March 18 2015 }

\begin{document}

\title{The Physical Conditions in a Pre Super Star Cluster Molecular
  Cloud in the Antennae Galaxies}

\author{K. E. Johnson\altaffilmark{1}, A. K. Leroy\altaffilmark{2},
  R. Indebetouw \altaffilmark{1,2}, C. L. Brogan\altaffilmark{2},
  B. C. Whitmore\altaffilmark{3}, J. Hibbard\altaffilmark{2},
  K. Sheth\altaffilmark{2}, A. Evans\altaffilmark{1,2}}

\altaffiltext{1}{Department of Astronomy, University of Virginia,
    Charlottesville, VA 22904-4325 {\textit{e-mail:}} kej7a@virginia.edu}
\altaffiltext{2}{National Radio Astronomy Observatory, 520 Edgemont Road,
  Charlottesville, VA 22903}
\altaffiltext{3}{Space Telescope Science Institute, 3700 San Martin Drive, Baltimore, MD 21218}

\begin{abstract}

  We present an analysis of the physical conditions in an extreme
  molecular cloud in the Antennae merging galaxies.  This cloud has
  properties consistant with those required to form a globular
  cluster.  We have obtained ALMA CO and 870$\mu$m observations of the
  Antennae galaxy system with $\sim 0''.5$ resolution.  This cloud
  stands out in the data with a radius of $\lesssim 24$~pc and mass of
  $>5\times 10^6$~M$_\odot$.  The cloud appears capable of forming a
  globular cluster, but the lack of associated thermal radio emission
  indicates that star formation has not yet altered the environment.
  The lack of thermal radio emission places the cloud in an early
  stage of evolution, which we expect to be short-lived ($\lesssim
  1$~Myr) and thus rare.  Given its mass and kinetic energy, for the
  cloud to be confined (as its appearance strongly suggests) it must
  be subject to an external pressure of P/$k_B \gtrsim
  10^8$~K~cm$^{-3}$ -- 10,000 times higher than typical interstellar
  pressure.  This would support theories that high pressures are
  required to form globular clusters and may explain why extreme
  environments like the Antennae are preferred environments for
  generating such objects.  Given the cloud temperature of $\sim
  25$~K, the internal pressure must be dominated by non-thermal
  processes, most likely turbulence.  We expect the molecular cloud to
  collapse and begin star formation in $\lesssim 1$~Myr.

\end{abstract}


\section{Introduction}

\subsection{Globular Cluster Formation}
Globular clusters are among the most ancient objects in the universe,
often with ages $>12$~Gyr \citep{bolte95, carretta00} and are common
around massive galaxies in the universe today \citep{harris13}.  The
present-day abundance of globular clusters is remarkable given that
that the fraction expected to survive $\sim 10$~Gyr is extremely
small, potentially lower than 1\% \citep{fall01, whitmore07}.  Thus, this
extreme type of star formation may have been a critical mode in the
early evolution of today's massive galaxies.

Initial theories about globular cluster formation suggested that these
objects were among the first to gravitationally collapse in the early
universe \citep{peebles68}.  Subsequent work, particularly after the
launch of the Hubble Space Telescope, has demonstrated that clusters
with extreme stellar densities often exceeding $\sim
10^4$~stars~pc$^{-3}$ \citep[e.g.][]{miocchi13} can still form in the
universe today \citep{oconnell94} -- the so-called ``super star
clusters'' (SSCs).  Since that time, numerous studies have indicated
that the properties of SSCs are consistent with those expected of
young globular clusters \citep[e.g.][]{mclaughlin08}.

A number of physical processes contribute to the destruction of clusters,
including two-body relaxation, stellar mass loss and feedback,
compressive and tidal shocks as clusters orbit their host galaxy, and
tidal truncation.  The extent to which each of these processes act on
a specific cluster will depend on a variety of factors, including the
orbital properties of the cluster \citep{gnedin97}.  Indeed, the extent to
which cluster disruption is mass-dependent is still debated
\citep{fall09, bastian12}.  For unresolved clusters in galaxies outside
the local group, there is typically limited (if any) dynamical
information, and for clusters younger than a few~Myr, little
dynamical evolution will have taken place.  For all of these reasons,
it is not possible to say whether any particular SSC will survive for
a Hubble time.  

While there is no generally accepted definition of ``Super Star
Cluster'', here we adopt a definition based on a cluster having
the {\it potential} to evolve into a globular cluster, regardless of
whether or not it actually will do so over the following $\sim$~10~Gyr.
This requirement results in both mass and radius limits on the range
of objects that can be considered as SSCs.  Specifically, most
present-day globular clusters have half-light radii of $< 10$~pc
\citep[although some have radii as large as $\sim
15$~pc,][]{vandenbergh91}, and stellar masses of $\gtrsim
10^5$~M$_{\odot}$ \citep{harris94}.  In addition, these clusters are
expected to lose $\gtrsim 1/2$ of their mass due to dynamical effects
over ~10$^{10}$ years \citep{mclaughlin08}, which suggests that to be
a globular cluster progenitor, a young star cluster should have a mass
of $\gtrsim 2\times 10^5$~M$_\odot$.  If star formation efficiency is
$\sim 20-50$\% \citep{ashman01, kroupa01}, the initial molecular core
from which the cluster is formed must have a mass of $\gtrsim
10^6$~M$_\odot$.

Optical techniques have been able to probe the evolution
of SSCs (and presumably some future globular clusters) to ages as
young as a few million years.  Before this time, the clusters can be
significantly shrouded by their birth material, limiting the
usefulness of optical observations.  Beginning in the late 1990's
efforts began to observe SSC evolution at even earlier ages
($\lesssim$ a few million years) by using radio observations to detect
the free-free emission from the ionized gas around the cluster and
internal to the cluster's dust cocoon \citep{turner98, kj99, turner00,
  beck00, johnson01, johnson02, johnson03, beck04, turner04,
  johnson04, reines08, johnson09, tsai09, aversa11, kepley14}.  We
refer to these objects as ``natal'' SSCs, meaning that the clusters
themselves have already formed, but they have not yet emerged from
their birth material.  Studies of natal clusters were able to place
constraints on the relative lifetime of this enshrouded phase of SSC
evolution to $\lesssim$~a million years and the gas density of the
ionized hydrogen $n_e > 10^3$~cm$^{-3}$\citep{johnson03}.  A
large number of subsequent studies have now identified additional
compact thermal radio sources in a number of galaxies, although their
low detection rate supports their relatively short lifetime
\citep{tsai09, aversa11}.

However, determining the physical conditions that give rise to SSCs
(and their surviving descendants -- globular clusters) has been mired
in the fundamental difficulty that once an SSC is present in the
molecular cloud, it will dramatically alter it.  Thus in order to
observationally probe the conditions capable of creating an SSC
requires not only identifying molecular clouds that are compact (radii
$\lesssim 25$~pc, see Section~\ref{cloud_size}) and massive ($\gtrsim
10^6$~M$_{\odot}$), but also for which massive stars have not begun to
disrupt the environment.  Efforts to observe the actual formation of
SSCs -- before the star clusters have formed -- requires high spatial
resolution millimeter observations to determine the physical
properties of the material from which the clusters will form.  Such
work has largely been stymied by the available observing facilities
and limited to only the most nearby galaxies.

One example of a relatively nearby starburst system in which some
progress has been made is M82.  At 3.6~Mpc \citep{freedman94},
relatively good linear resolution was achievable even before ALMA.
This system was observed using the Owens Valley Radio Observatory
(OVRO) in CO(2-1) with a linear resolution of 17~pc \cite{keto05}.
While the compact molecular clouds observed in M82 are likely to be
associated with early SSC evolution, multiwavelength observations
suggest that these clouds have already begun star formation
\citep{keto05}, and have therefore disrupted their birth
environment\footnote{A possible exception to this is a CO cloud
  located at 09h55m54.5s +69d40’50”, however the properties of this
  cloud are not provided by \citet{keto05}.}.

This paper reports results from an ALMA Early Science project
  studying the Antennae galaxies. In a high resolution survey of CO
  emission from the Antennae \citep{whitmore14} the most immediately
  striking feature was a compact, high line width cloud with little
  associated star formation. This is coincident with a source
  identified by \citet{herrera11,herrera12} as a potential proto-SSC
  using H$_2$ and earlier, lower resolution ALMA data.  The strong
  compact H$_2$ emission appears to be due to warm (1700-2300~K)
  shocked gas.  However, the size of the cold molecular component
  could only be constrained to $\lesssim 100$~pc, which precluded a
  conclusive identification.

The present paper characterizes this source, which we consider among
the best candidates for a proto-SSC, and lays out the evidence for and
against the source's eventual evolution into a SSC or GC.

\begin{figure*}
\begin{center}
\includegraphics[width=7in]{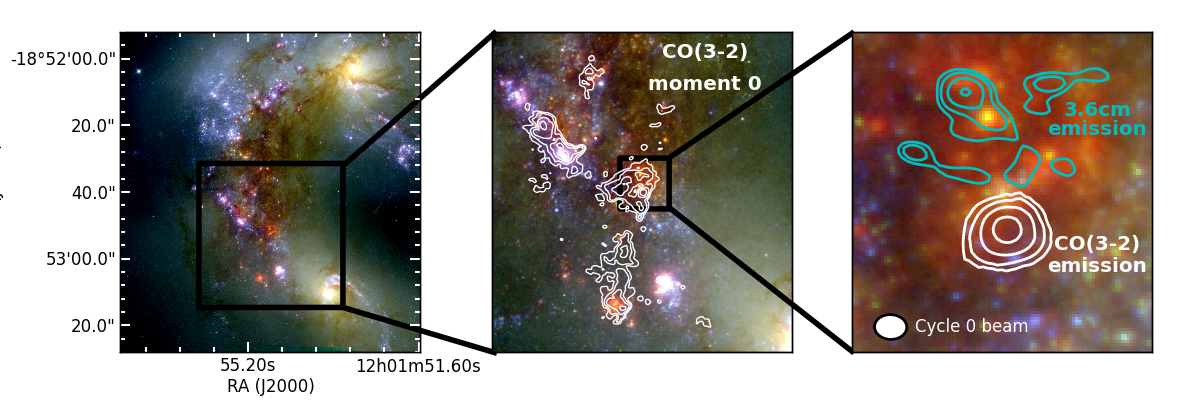}
\caption{The candidate proto super star cluster molecular cloud is located in
  the ``Overlap'' region of the Antennae merging galaxies.  (left) HST
  color image of the central area of the Antennae Overlap region (red
  = Pa$\alpha$, green = F814W ($\sim$ I-band), blue = F435W
  ($\sim$B-band)), (middle) A magnified view of the ``overlap'' region
  overlaid with contours (2, 4, 8, and 16 Jy beam$^{-1}$ km~s$^{-1}$)
  from the ALMA Cycle 0 CO(3-2) moment 0 map. (right) A zoomed-in view
  of the region surrounding the candidate proto super star
  cluster molecular cloud overlaid with
  both CO(3-2) contours of the molecular cloud after extraction from
  the data cube (0.4, 0.8, 1.6, 3.2 Jy beam$^{-1}$ km s$^{-1}$) and
  3.6cm radio emission (-1, 4, 5, 7, 10$\sigma$, $\sigma = 3.8\times
  10^{-2}$~mJy~beam$^{-1}$). The synthesized beam, shown in the
    right panel, has a size of $0.''56 \times
    0.''43$. \label{plot_zoom_radio}}
\end{center}
\end{figure*}

\begin{deluxetable}{cccc}
\tablecaption{Interferometric CO Observations of the Antennae \label{previous_obs}}
\tablehead{
\colhead{Facility} & \colhead{Transition} & \colhead{Beam} & \colhead{Reference}
}
\startdata
CMA      & 12CO1-0& $4.91''\times 3.15''$ & \cite{wilson00} \\
SMA      & 12CO3-2& $1.42''\times 1.12''$ & \cite{ueda12} \\
SMA+PdBI & 12CO2-1& $3.3'' \times 1.5''$ & \cite{wei12}\\
ALMA-SV  & 12CO2-1& $1.68''\times 0.85''$ & \cite{espada12}\\
ALMA-SV  & 12CO3-2& $1.05''\times 0.60''$ & \cite{herrera12}\\
ALMA-Cyc0  & 12CO3-2& $0.56''\times 0.43''$ & {\small this paper} \\ 
\enddata
\end{deluxetable}

\section{ALMA Observations}
We obtained ALMA observations of the Antennae system in CO(3-2) and
870$\;\mu$m continuum with the goal of probing the conditions of
cluster formation and early evolution; data calibration is discussed
in detail in the overview paper \citep{whitmore14}.  Briefly, the
observations consisted of a 13-point mosaic, and were carried out in
the ``extended'' configuration, with a maximum baseline of $\sim
400$~m and 5~km~s$^{-1}$ spectral channels.  The resulting rms was
  determined using line-free channels and found to be 3.3~mJy/beam.
With an angular resolution FWHM of $0.''56 \times 0.''43$ ($59 \times
45$ pc), these observations are well-matched to the expected diameter
of the precursor giant molecular clouds of $\lesssim 50$~pc (or a
radius of 25~pc, see Section~\ref{cloud_size}).  For this paper, we
also recalibrated and reimaged the SV data, as well as compared it to
previous results to check for consistency.  The SV CO(2-1) data used
  in this paper was taken with a beam size of $1.68''\times 0.85''$.
  The rms of the SV CO(2-1) observations was also measured using
  line-free channels and found to be 6.5~mJy/beam.

\begin{figure*}
\begin{center}
\includegraphics[width=7in]{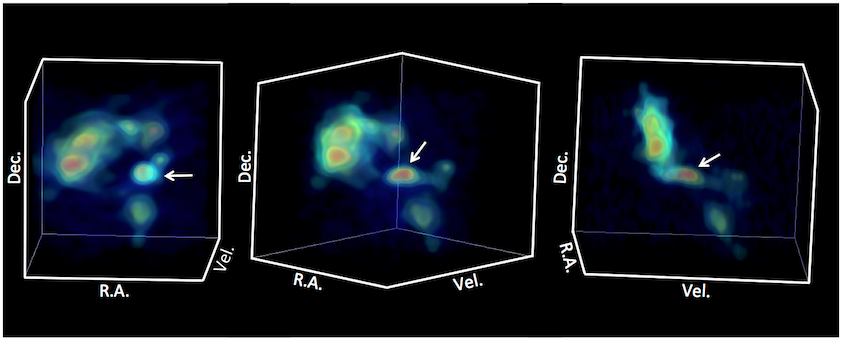}
\caption{A 3-D view of the data cube illustrates the spatial and
  velocity structure of the entire SGMC2 region.  The candidate
  proto-super star cluster molecular cloud is indicated by the arrow,
  and the secondary cloud is spatially offset slightly to the
  north-west; the primary cloud stands out in the data cube due to its
  round shape, compact size, and larger velocity dispersion (it is
  visibly extended in velocity space, in contract to other clouds in
  the cube, including the secondary cloud which is offset to higher
  velocity).  The spatial dimensions of the cube are $9.''6 \times
  9.''6$, and the velocity axis spans
  1300-1800~km~s$^{-1}$. \label{3D}}
\end{center}
\end{figure*}

The ALMA observations enable the study of a compact and luminous
source in the CO(3-2) data cube (see Figure~\ref{plot_zoom_radio}).
This cloud is part of the super giant molecular cloud complex known as
SGMC2 \citep{wilson00}; the full 3-D cube of SGMC2 is shown in
Figure~\ref{3D}.  The Antennae galaxy system was previously observed
in both CO(2-1) and CO(3-2) by ALMA as part of the ``science
verification'' (SV) process, which has already resulted in
publications \citep[see Table~\ref{previous_obs},][]{espada12,
  herrera12}.  

The specific source discussed here was singled out in the
lower-resolution ALMA science verification data, with the CO(3-2)
emission being coincident with strong H$_2$ emission -- potentially
indicating shocks due to infalling gas \citep{herrera12}. Even with
lower resolution data ($\sim 100$~pc), it was speculated that this
region might contain an SSC in the early stages of its evolution
\citep{herrera12}.  However, the spatial resolution of the SV CO(3-2)
data is $\sim 100$~pc, or roughly twice that of the Cycle~0 data
presented here.  It is clear that in the SV data, the molecular cloud
that is the subject of this paper is not resolved and is blended with
other molecular material in the vicinity.

\begin{deluxetable*}{cccccc}
\tablecaption{Measured Properties of the Molecular
  Cloud \label{properties}}
\tablehead{
\colhead{R.A.} & \colhead{Dec.} & \colhead{V$_{LSR}$} &
\colhead{S$_{CO3-2}$} &  \colhead{$\sigma_V$} & 
\colhead{Size}\\
\colhead{(J2000)} & \colhead{(J2000)} & \colhead{(km s$^{-1}$)} & 
\colhead{(Jy km s$^{-1}$)} & \colhead{(km s$^{-1}$)} & \colhead{FWHM (arcsec$^2$)}
}
\startdata
12:01:54.73 & -18:52:53.2 & 1524$\pm 3$ & 52$\pm 5$ & 49$\pm 3$ &
$0.66\pm0.12 \times 0.55\pm0.07$\\
\enddata
\tablecomments{Properties of the clouds were measured both with the
  CPROPS program \cite{rosolowsky06}, and with Gaussian fitting (FWHM
  $= 2.35\sigma$).  The quoted uncertainties reflect empirically
  determined variations in these values for different fitting
  attempts. The synthesized beam has $\it not$ been deconvolved in
  this table to strictly report measured properties.  The deconvolved
  size is reported in Table~\ref{inferred_properties}.}
\end{deluxetable*}

\subsection{Cloud Analysis}
\begin{figure}
\begin{center}
\includegraphics[width=1.0\columnwidth]{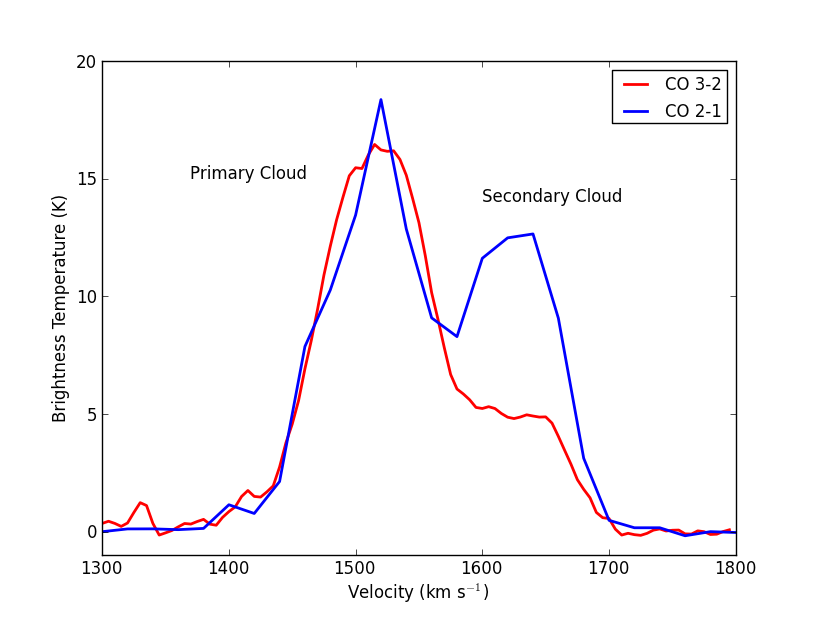}
\caption{The CO(3-2) and CO(1-0) spectra of a $\sim 1''\times 1''$
  region around the candidate proto super star cluster molecular cloud
  taken from the full data cube (i.e. the primary cloud has not been
  extracted).  The line emission indicates that there are two
  components along the line-of-sight that have different
  temperatures. Spectral profile of the CO(3-2) line from current work
  (ALMA Cycle~0 observations) and the CO(2-1) line from ALMA science
  verification \citep{espada12}.  The CO(3-2) data is convolved to the
  CO(2-1) beam and both data sets are corrected for beam dilution.
  There is clearly more than one velocity component; in these
  convolved data, the secondary source was fit by a Gaussian and
  subtracted from the spectra.  The CO(3-2)/CO(2-1) ratio is
  dramatically different in the two velocity
  components. \label{line_profile}}
\end{center}
\end{figure}

\begin{figure}
\begin{center}
\includegraphics[width=1.\columnwidth]{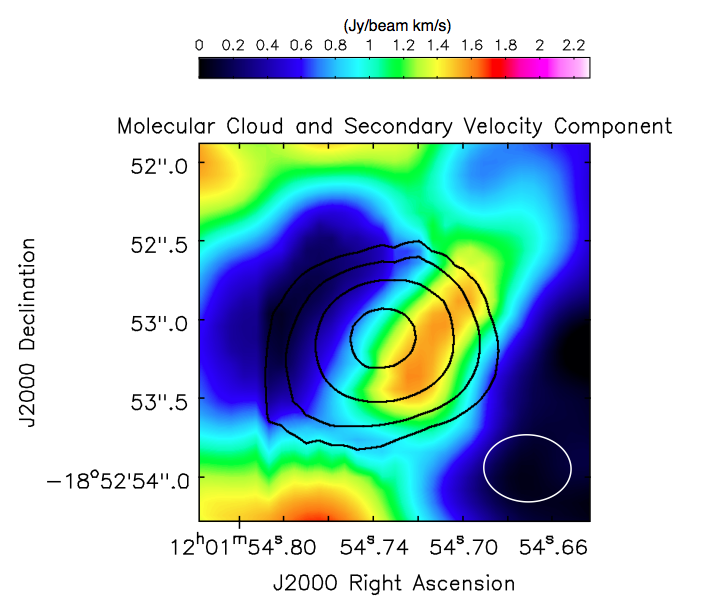}
\caption{
 These observations allow us to disentangle the molecular
  cloud associated with the secondary velocity component.  Contours of
  the CO(3-2) moment~0 map of the molecular cloud extracted from the
  3D data cube (0.4, 0.8, 1.6, 3.2 Jy beam$^{-1}$~km~s$^{-1}$)
  overlaid on the color moment~0 map of the data cube with the primary
  cloud extracted.  The cloud corresponding to the secondary velocity
  component can be seen in spatial projection with the primary
  cloud.  \label{extracted_cloud}}
\end{center}
\end{figure}

Determining the properties of this cloud requires that it first be
isolated from other emission in the region.  As shown in
Figures~\ref{line_profile} and \ref{extracted_cloud}, there is a
redshifted secondary velocity component along this line of sight, 
  that must be deconvolved from the primary source before analysis.
We extract the primary cloud from the velocity cube for further
analysis by creating a sub-cube around the cloud with dimensions
  of $\sim 1''\times 1'' \times 300$~km~s$^{-1}$.  The integrated
intensity contours of the extracted primary cloud are shown in
Figure~\ref{plot_zoom_radio}, and the observed properties are listed
in Table~\ref{properties}.  We determine the cloud size by
deconvolving the synthesized beam from the extracted source, which
yields a half-light radius for the cloud of $\lesssim 24\pm 3$~pc
($\lesssim 0.23''$).  The derived properties of the cloud are given in
Table~\ref{properties}.  As a sanity check, the properties of the
primary cloud were also determined using the CPROPS program on the
entire data cube (not exclusively the sub-cube) \citep{rosolowsky06},
and the resulting parameters agree to within the uncertainties. 
  Thus, by-hand measurement of the half-light size, automated Gaussian
  fits, and moment based measurements all yield roughly consistent
  sizes for our cloud. As this is a marginally resolved object with a
  clearly measurable line width methodological uncertainties do not
  overwhelm any of our conclusions.  Throughout this text, we
  refer to the properties of the extracted primary cloud only, unless
  noted otherwise.

\subsection{Relative Astrometric Solutions}
A comparison between the CO(3-2) emission and data at other
wavelengths, requires an understanding of the relative astrometric
accuracy.  Based on the phase stability of the ALMA observations, we
estimate the absolute astrometric accuracy to be better than $\sim
0.2''$.  Centimeter observations from the VLA have an astrometric
accuracy better than $\sim 0.1''$ \citep{brogan10}, and
therefore the 3.6~cm and CO(3-2) observations have a relative
precision of better than the synthesized beam of the ALMA data, and we
consider them to be astrometrically matched.  We also register the
astrometry of archival Hubble Space Telescope observations shown in
Figure~\ref{plot_zoom_radio} by matching the Pa${\alpha}$ emission
throughout the Antennae system to common features in the 3.6~cm
emission.

\section{Results}

\subsection{Cloud Temperature and Optical Depth}
We constrain the temperature and optical depth of the cloud using the
CO(3-2) and CO(2-1) emission. We retrieved, recalibrated, and reimaged
CO(2-1) observations from ALMA's science verification
period, shown over-plotted in Figure~\ref{line_profile}.
The CO(3-2) observations were convolved to the synthesized beam of the
CO(2-1) observations and corrected for beam dilution using the Cycle~0
CO(3-2) source size, resulting in peak brightness temperatures of T$_{3-2}
= 17\pm3$~K and T$_{2-1} = 18\pm3$~K.  The largest angular scale to
which the CO(3-2) observations are sensitive is $\sim 6''$, and
therefore we do not expect that any flux is resolved out on the size
scales of interest here. The secondary component has a significantly lower
CO(3-2)/CO(2-1) ratio, indicating much cooler gas than the primary
cloud.  

RADEX non-LTE modeling \citep{vandertak07} was used to analyze the
CO(3-2) and CO(2-1) emission.  The line intensities and their
  ratio were compared to a grid of RADEX models covering a range of
values for kinetic temperature, CO column density, and H$_2$ volume
density.  The best-fit values result from a chi-squared minimization.
Since the source is marginally resolved in CO(3-2), for CO(3-2)
  we set the beam filling fraction to 1, and for CO(2-1) we set it to
  the dilution factor, or the ratio of the CO(3-2) size to the CO(2-1)
beamsize.
The RADEX models indicate that these transitions are optically thick
-- the best fitting depth is $\tau\sim 3.5 \pm 0.5$, but the data do
not rule out significantly higher values.  The lines appear to be
close to thermalized, with an excitation temperature within a degree
of the kinetic temperature of 25$^{+10}_{-2}$~K; this temperature is
on the upper end of the range of those found for dense molecular
clouds in the Milky Way \citep{shirley13}.  However, there is a
degeneracy between the inferred temperature and density of the cloud,
and the cloud could be warmer for densities $\lesssim 6\times
10^4$~cm$^{-3}$.  In other words, the observed brightness could
  also be reproduced with a large column of subthermally excited, warm,
  relatively diffuse gas.

\begin{deluxetable*}{ccccccccc}[!b]
\tabletypesize{\footnotesize}
\tablecaption{Derived Properties of the Molecular
  Cloud \label{inferred_properties}}
\tablehead{
\colhead{Deconv.} & \colhead{Half Light} & \colhead{M$_{virial}$} & \colhead{M$_{X_{CO}}$} &
\colhead{M$_{RADEX}$} & \colhead{M$_{Cont.}$} & \colhead{T$_{Kin}$} & \colhead{$\rho$}\\
\colhead{FWHM (pc$^2$)} & \colhead{R (pc$^2$)} & \colhead{($10^6$~M$_{\odot}$)} & \colhead{($10^6$~M$_{\odot}$)} &
\colhead{($10^6$~M$_{\odot}$)} & \colhead{($10^6$~M$_{\odot}$)} & \colhead{(K)} & \colhead{(g cm$^{-3}$)}
}
\startdata
$< 53\times 41$ & $< 27\times 21^c$ & $29-85$ & $3.3-15$ & $> 2.8$ &
$3.1-7.4$ & $23 - 35$ &
$2 - 20 \times 10^{-21}$\\
\enddata
\tablecomments{ The FWHM as measured after deconvolving the
  synthesized beam and adopting a distance of 21.5~Mpc.  The half
  light radius resulting from deconvolution assuming a Gaussian
  distribution of light.  The cloud is consistent with either being
  marginally resolved or a point source at this resolution. }
\end{deluxetable*}

The temperature inferred here for the CO cloud is roughly 100$\times$
less than that inferred for the compact H$_2$ emission observed in
this region of 1,700-2,300~K \citep{herrera11}.  In addition, the
H$_2$ FWHM line-width of $\sim 150$~km~s$^{-1}$ \cite
{herrera11} is significantly higher than the  FWHM line-width of the
CO emission measured here of 115~km~s$^{-1}$ ($\sigma$ =
49~km~s$^{-1}$).  Therefore we infer that the origin of the CO and
H$_2$ emission may not be identical.  We speculate that the H$_2$
emission has a low filling factor, sampling only the most strongly
shocked regions in and/or around the cloud.

\subsection{Cloud Mass \label{mass}}
The mass of the cloud is estimated using four different methods, each
subject to different caveats.  First, given the source size, velocity
dispersion of $\sigma_v = 49\pm 6$~km~s$^{-1}$, and assuming an isothermal
  sphere we calculate the virial mass to be in the range of
M$_{vir} = 2.9-8.5\times 10^7$~M$_{\odot}$, which is consistent with
the virial mass estimated from ALMA SV observations of $\sim 5\times
10^{7}$~M$_{\odot}$ \citep{herrera12}.  This mass estimate will only
be valid if the cloud is in gravitational virial equilibrium; any
additional velocity in the cloud will result in this method
overestimating the mass.  Given the complex dynamical structure of
SGMCs, we treat the estimated virial mass as an {\it upper} limit.

For the second method we use the RADEX models of the CO(3-2) and
CO(2-1) observations to determine  the $\chi^2$ best fit column
density of N$_{CO} \gtrsim 5\times 10^{18}$~cm$^{-2}$.  If we assume
an abundance ratio of $n_{H2}/n_{CO} = 2 \times 10^4$
\citep{blake87,wilson92}, this column density of CO corresponds to an
H$_2$ mass of M$_{non-LTE} = 2.8 \times 10^6$~M$_{\odot}$.  However,
the $\chi^2$ values are shallow toward higher values of N$_{CO}$,
  and only weakly constrain the upper bound. The models also indicate
that the CO(3-2) has an optical depth of $\tau \gtrsim 3$, and
therefore this mass estimate is a {\it lower} limit.

The third method we employ assumes a conversion factor, X$_{CO}$.
  Values for X$_{CO}$ in starbursts are known to vary by at least a
  factor of four \citep{bolatto13}, and thus the mass estimated using
  X$_{CO}$ should be regarded as uncertain by a corresponding factor.
  Here we adopt a ``starburst'' CO-to-H$_2$ conversion factor X$_{CO}
  = 0.5\times 10^{20}$~cm$^{-2}$(K~km~s$^{-1}$)$^{-1}$
  \citep{bolatto13}. If we were to adopt a ``standard'' X$_{CO} =
  2\times 10^{20}$~cm$^{-2}$(K~km~s$^{-1}$)$^{-1}$, the resulting
  cloud mass would be a factor of four larger.  We further assume that
CO(3-2) is thermalized with respect to CO(1-0) given the brightness
temperatures of T$_{3-2} = 17\pm3$~K and T$_{2-1} = 18\pm3$~K ; if the
lines are not thermalized and CO(3-2) is relatively underpopulated,
this mass estimate will be low.  This method results in M$_{X_{CO}} =
3.3-15\times 10^6$~M$_{\odot}$.

\begin{figure}
\includegraphics[width=1.\columnwidth]{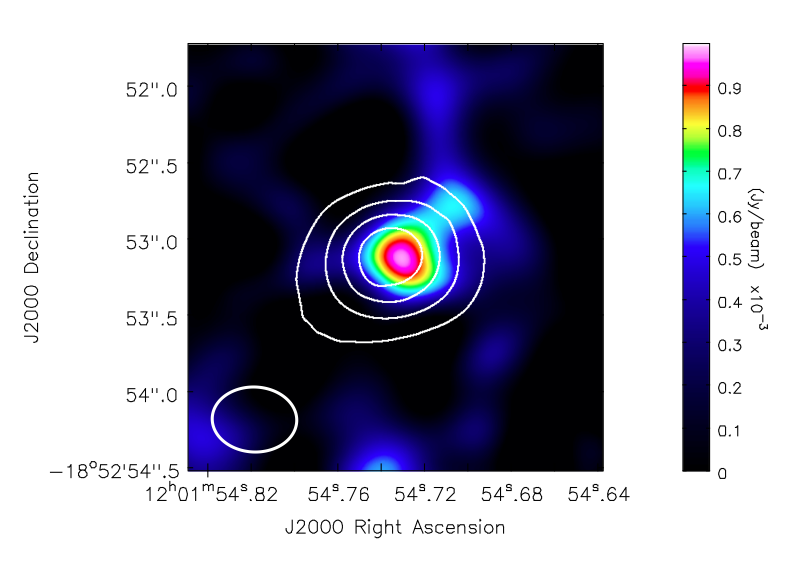}
\caption{ Detecting the continuum emission
  associated with this source allows us to determine its dust mass.
  Contours of the CO(3-2) moment 0 map of the primary cloud overlaid
  on a color-scale image of the 870~$\mu$m continuum. \label{continuum}}
\end{figure}

The last method estimates the dust mass from the continuum at
870~$\mu$m.  The continuum is detected at the 5.4~$\sigma$ level (see
Figure~\ref{continuum}), with a peak brightness of $9.8\pm3.4 \times
10^{-4}$~Jy~beam$^{-1}$.  Assuming a dust emissivity of $\kappa = 0.9
\pm 0.13$~cm$^{2}$~g$^{-1}$ \citep{wilson08}, optically-thin
continuum, and a gas-to-dust ratio of 120$\pm 28$, and dust
temperature of 20~K \citep{wilson08} this method results in a mass of
M$_{cont} = 5\pm3 \times 10^6$~M$_{\odot}$.  Based on the range of
mass values determined above, we adopt a cloud mass of M$= 0.3 - 1.5
\times 10^7$~M$_{\odot}$.  We note that the virial mass that would be
inferred for this cloud appears to be a factor of 5-10$\times$ too
high.

\begin{figure*}
  \begin{center}
\includegraphics[width=1.5\columnwidth]{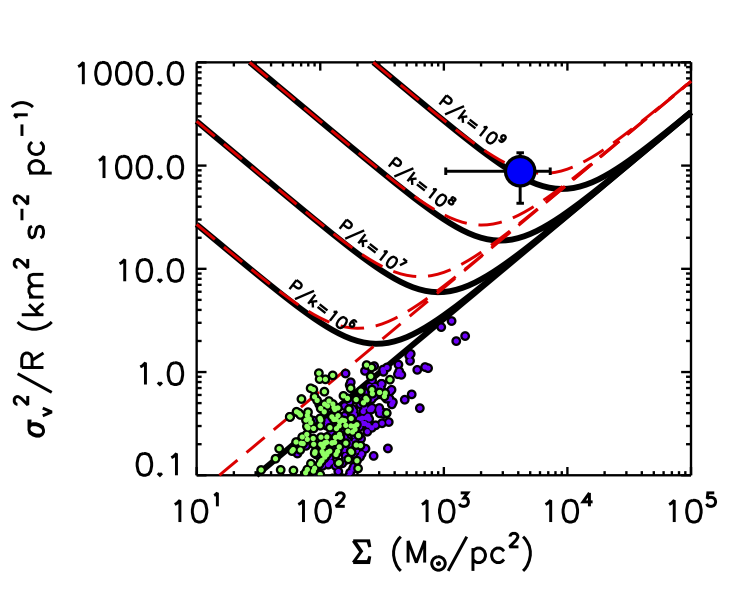}
\caption{The measured properties of the primary molecular cloud
  indicate that it may be subject to pressures of $\sim$P/$k_B =
  10^9$~K~cm$^{-3}$.  This plot shows the relationship between surface
  density and the size-line width coefficient with the candidate
  proto-super star cluster cloud plotted as the blue point (error bars
  represent 1$\sigma$). Lines corresponding to pure gravitational
  virial equilibrium (no external pressure) along with corresponding
  line for equilibrium with external pressures of P$/k_B = 10^6, 10^7,
  10^8$, and $10^9$ ~K~cm$^{-3}$ are shown in black, corresponding
  conditions for clouds undergoing free-fall collapse are shown as red
  dashed lines \citet{field11}. For comparison, data points are shown
  from molecular cloud survey in nearby galaxies as light green points
  \citep{meyer13}, and Galactic clouds are shown as purple points
  \citep{heyer09}. \label{pressure}}
\end{center}
\end{figure*}

Given the estimated mass and size of this cloud, the resulting gas
volume density is $\rho \gtrsim 100$~M$_{\odot}$~pc$^{-3}$. While
there are molecular clouds found in the Milky Way with masses of $\sim
10^6$~M$_{\odot}$, their radii are $~2-4\times$ larger, resulting in
significantly lower surface densities.  Similarly massive clouds have
also been identified in other nearby galaxies \citep{bolatto08,
  meyer13}; their surface densities are also far lower (see
Figure~\ref{pressure}).

We estimate the mass of the stellar cluster that will potentially
result from this molecular core by assuming a star formation
efficiency (SFE, fraction of mass turned into stars over the lifetime
of a cloud).  The net efficiency can vary wildly, ranging from a
  few percent in the Milky Way \citep[e.g.][]{lada03}, to $>50\%$ in
  cluster-forming cores by basic boundedness arguments.  If we adopt a
  SFE typical for clusters of $\epsilon = M_{stars}/(M_{stars} +
M_{gas})$ of $\sim 20\% - 50\%$ \citep{ashman01, kroupa01}, an initial
cloud mass of $5\times 10^6 - 10^7$~M$_{\odot}$ would result in a
cluster with a mass of $M_{star} = 1-5 \times 10^6$~M$_{\odot}$.  This
would be among the most massive SSCs that have formed in the Antennae
if it forms a single cluster \citep{whitmore10}.  Even if the
  star formation efficiency were as low as $\sim 5\%$, the resulting
  cluster would have a mass $> 2 \times 10^5$~M$_{\odot}$, still in
  the regime of super star cluster masses.

\subsection{Constraining the Ionizing Flux Associated with the Cloud}
In order to assess the extent to which star formation may have already
affected the physical state of the ionized gas, we searched for
ionizing flux potentially coming from stars within the molecular
cloud.  Figure~\ref{plot_zoom_radio} shows the Pa$\alpha$ emission in
the region, and while there is diffuse emission associated with the
SGMC in general, there is no discrete source associated with the
molecular cloud in question.  However, it is possible that given the
embedded nature of this source, Pa$\alpha$ emission could suffer from
significant extinction, and thus we also utilize radio observations to
identify free-free emission that might be present.

We created two radio maps from archival 3.6~cm VLA observations
(proposal codes AN079, AP478, AS796, AA301); one with a synthesized
beam of $0.65''\times 0.42''$ to best match the beam of the ALMA
CO(3-2) observations and resolve out diffuse emission, and a second
with a synthesized beam of $1.12'' \times 0.85''$, which has greater
sensitivity to extended emission.  In the higher resolution map, there
is no discrete source coincident with the molecular cloud discussed
here, and the 5$\sigma$ detection threshold of the radio emission
corresponds to an ionizing flux of N$_{Lyc} \approx 6\times
10^{50}$~s$^{-1}$.  This is equivalent to $\sim 60$ O-type stars
\citep{vacca96}.  This limit is roughly three times lower than the
previous limit placed on possible ionizing flux in this region of
$2\times 10^{51}$~s$^{-1}$ using lower resolution 6~cm observations
\citep{herrera11}.  For comparison, the ionizing flux of 30~Dor
  has been estimated to be N$_{Lyc} \approx 4\times 10^{51}$~s$^{-1}$
  \citep{crowther98}.

In the lower resolution map, while there is no discrete source
coincident with the molecular cloud, there is diffuse emission in the
region.  Without velocity information for this diffuse emission,
  it is not possible to disentangle potential line-of-sight confusion
  in this complex region.  However, to estimate the possible
contribution from stars within this molecular cloud to the diffuse
emission, we first fit and subtract Gaussian profiles to the other
dominate sources in SGMC2.  After subtracting the emission due to
nearby sources, we measure the flux density due to diffuse emission in
the cloud aperture to be $0.03$~mJy, which corresponds to an ionizing
flux of $1.1\times 10^{51}$~s$^{-1}$, which corresponds to $\sim 100$
O-type stars \citep{vacca96}.  However, the diffuse morphology of this
emission is not consistent with it coming from a single compact
source.

Thus, while both the Pa${\alpha}$ and cm radio observations show
diffuse emission associated with the SGMC2 structure, there is no
discrete source associated with the compact cloud discussed here
(Figure~\ref{plot_zoom_radio}).  We constrain the possible ionizing
flux that could be due to embedded stars in this cloud to be $\lesssim
60$ O-type stars.  We can rule out an existing stellar cluster in this
molecular core $\gtrsim 10^4$~M$\odot$ \citep{leitherer99}, roughly
two orders of magnitude smaller than the anticipated cluster mass
(Section~\ref{mass}). Either star formation has not begun or it is so
deeply embedded that its ionizing radiation is confined by gas
continuing to accrete onto the protostars.

\subsection{Determination of Cloud Pressure}
With a surface density of $\sim 4 \times 10^3$~M$_{\odot}$~pc$^{-2}$
and size-linewidth coefficient of $\sigma^2/R =
90$~km$^2$~s$^{-2}$~pc, the cloud is not consistent with being in
either pure gravitational virial equilibrium or free-fall collapse
\citep[Figure~\ref{pressure},][]{heyer09}.  Nevertheless, the
morphology of the cloud indicates that gravity is playing a
significant role (round, compact, and bright -- making it stand out as
a singular object in the data cube), suggesting that it is not a
transient object.  As illustrated in Figure~\ref{pressure}, the
observed line width value can be explained if the cloud is subject to
external pressures of P/$k_B \sim 10^9$~K~cm$^{-3}$, roughly five
orders of magnitude higher than that typical in the interstellar
medium of a galaxy \citep{jenkins83}. This is consistent with
theoretical considerations that have argued SSC formation requires
extreme pressures (P/$k_B \gtrsim 10^8$~K~cm$^{-3}$) \citep{jog92,
  elmegreen97,ashman01}.  This high pressure is also in accord with
previous findings of compressive shocks in the overlap region
\citep{wei12}.

\subsection{Expected Proto Super Star Cluster Cloud Size \label{cloud_size}}
The expected physical size of a molecular cloud capable of forming an
SSC can be estimated based on virial theorem arguments.  Following
previous work \citep{elmegreen89}, the external cloud pressure $P_e$,
cloud mass $M$, cloud radius $r$, and velocity dispersion $\sigma_v$
can be related by

$P_e = \dfrac{3 \Pi M \sigma_v^2}{4 \pi r^3}$, 

\noindent where $\Pi$ is defined by $n_e = \Pi \langle n_e \rangle$,
and here we adopt $\Pi = 0.5$.  It has been estimated that SSC
formation requires internal pressures of $P_0/k\gtrsim
10^{8}\;$K$\;$cm$^{-3}$ \citep{elmegreen97}.  These high pressures
inhibit the dispersal of the natal material and achieve sufficiently
high star formation efficiencies in the cloud core.  If we adopt a
minimum mass of $M=10^6$~M$_{\odot}$ for a cloud capable of forming a
SSC, the resulting cloud radius is $r\sim25$~pc.  Likewise, for the
velocity dispersion of $\sigma_v = 49$~km~s$^{-1}$ observed for the
cloud discussed here and the apparent external pressure of
$P_0/k\gtrsim 10^{9}\;$K$\;$cm$^{-3}$ (see Figure~\ref{pressure}), the
expected radius is $r\sim 25$~pc, which is within the uncertainty of
the cloud half-light radius of $\lesssim 24 \pm 3$~pc extracted from
these observations.

\section{Discussion}

\subsection {On the Origin of the High Pressure Inferred for the Molecular Cloud}

If the cloud is confined, as we expect from such a strong
concentration of gas, then external pressure must play a key role.
This high external pressure could result from the weight of the
surrounding SGMC2 (if the SGMC is roughly in hydrostatic equilibrium)
and/or large scale compressive shocks.  The pressure generated by the
surrounding molecular material can be estimated using P$_G/k_B \approx
1.5$~cm$^{-3}$~K~(M$_{cloud}$/M$_{\odot}$)$^2$(r/pc)$^{-4}$
\citep{bertoldi92}.  The SGMC2 region has an estimated total mass of
$\sim 4\times 10^8$~M$_{\odot}$ \citep{wilson00} and radius of $\sim
400$~pc, resulting in a pressure from the overlying molecular material
of P/$k_B\sim 10^7$~K~cm$^{-3}$, which is at least an order of
magnitude less than the internal pressure inferred for this cloud.

Given that this cloud is not only in the ``overlap'' region of the
Antennae, but also appears to be at the nexus of two filaments of
CO(3-2) emission that are suggestive of colliding flows
\citep{whitmore14}, a significant amount of external pressure could
also be generated by ram pressure.  This interpretation is consistent
with previous work indicating strong H$_2$ line emission and an abrupt
velocity gradient across this region \citep{herrera11,herrera12}.
However, the morphology of the cloud suggests an isotropic source
  of pressure, which is in tension with a ram pressure origin.

\subsection{Timescales for Cloud Evolution and Star
  Formation \label{timescales}}

The relevant timescales for this cloud to evolve are driven by the
free-fall time and the crossing time.  The compact size and
marginally-resolved round morphology suggest that self-gravity has had
a significant role in shaping the source, although that is not
possible to conclusively demonstrate with the data in hand.  The cloud
must be largely supported by turbulence: given the inferred density of
the cloud (n$\sim 10^3$~cm$^{-3}$), if the pressure were entirely
thermal, it would require a gas temperature of $\sim 10^5$~K, which is
not reasonable or consistent with these observations that indicate
$T\sim 25$~K.  We conclude that this cloud is most likely supported by
turbulence.  This conclusion is also supported by the virial mass
being nearly an order of magnitude larger than the mass inferred from
the dust continuum, indicating significant internal motion
contributing to the observed line width.

Thus, we adopt the turbulent crossing time as the appropriate
timescale for the evolution of this cloud and estimate it as
$t_{cr} \sim D/\sigma_V \sim 1$~Myr, where $D$ is the diameter of the
region.   If self-gravity is {\it not} important, the cloud will
  disperse on this timescale -- the turbulence will dissipate on this
  timescale in any case.  If self-gravity is important (as argued
  above), the cloud will collapse on this timescale.  The
free-fall time of the cloud can be estimated as
$t_{ff}=(3\pi/32~G~\rho)^{1/2} \approx 8 \times 10^5$~years.  Given
the strong associated H$_2$ emission \citep{herrera12} and line-width,
which indicates internal velocities higher than virial equilibrium, it
is plausible that this cloud has already begun free-fall collapse, in
which case we are witnessing a very short-lived stage of cluster
evolution.  Given these arguments, we expect this cloud to collapse on
timescales $\lesssim 1$~Myr.

\subsection{Expected Number of Proto-SSC Molecular Clouds}

Given that this phase of SSC formation is expected to be extremely
short-lived ($\lesssim 10^5 - 10^6$~yr), it is not surprising that
this is {\it the only} clear example that we have found to date of a
compact cloud {\it without} associated star formation that is
sufficiently massive to form an SSC.  Within $\sim 1$~Myr, we expect
that this cloud will be associated with star formation, similar to the
clouds observed in M82 by \citet{keto05}.

The estimated current star formation rate (SFR) for the Antennae is
$\sim 7 - 20$~M$_{\odot}$~yr$^{-1}$ \citep{zhang01, brandl09}.  If cluster
formation follows a power-law distribution of $dN/dM \propto M^{-2}$
\citep{zhang99} with lower and upper masses of $10^2$ and
$10^7$~M$_{\odot}$, we expect $\sim 20$\% of the stellar mass to be
formed in clusters with M$> 10^6$~M$_{\odot}$.  Thus we expect only a few SSCs
masses of $>10^6$~M$_{\odot}$ to form every $\sim 5\times
10^5$~years.  Optical studies indicate that there are six SSCs with
ages $<10^7$ years and masses
$>10^6$~M$_{\odot}$ \citep{whitmore10}.  This suggests that a
massive cluster is formed every
$<$10$^7$/6~$\approx$~1.7$\times$10$^6$ years, in agreement with the
estimate derived above from the total SFR.   The predicted number of
pre-stellar molecular clouds capable of forming an SSC with mass $>
10^6$~M$_\odot$ can be generalized as, 
\begin{equation}
N_{\rm SSC-GMC}\,\simeq\,0.2 \times t_{collapse}\left(\frac{\rm SFR{(M_\odot yr^{-1})}}{\rm 10^6
    M_{\odot} }\right)
\end{equation}
where t$_{\rm collapse}$ is the timescale for cloud collapse (see
Section~\ref{timescales}).  Given a dynamical timescale for this
cloud of $\sim 0.5-1$~Myr, we expect that if we observed the Antennae
system at any point in its recent star forming history ($\sim
10^{7-8}$~yr), we would find at most one cloud of this evolutionary state
and mass.

\subsection{Implications for Globular Cluster Formation}

The physical properties of this cloud can provide insight into a mode
of star formation that may have been dominant in the earlier universe,
when globular clusters were formed prolifically. In particular, the
pressure of P/$k_B\gtrsim 10^8$~K~cm$^{-3}$ supports the hypothesis
that such high pressures are necessary to form SSCs \citep{ashman01,
  elmegreen02}.  Based on what appears to be a nearly universal
power-law distribution of cluster masses, numerous studies have
suggested that the formation of SSCs is statistical in nature,
resulting from a ``size of sample'' -- galaxies with higher star
formation rates will form more clusters overall, and the formation of
SSCs results from populating the tail of the mass distribution
\citep{fall12}.  Based on these ALMA observations, we suggest an
alternate interpretation: in a turbulent interstellar medium, 
the pressure distribution also has a power-law form, and the
properties of clusters that form track the pressure distribution.
Thus, while galaxies present a power-law distribution of cluster
masses, only regions with sufficiently high pressure will be
able to form the most massive SSCs. If pressures P/$k_B\gtrsim
10^8$~K~cm$^{-3}$ are indeed required to form SSCs (and the surviving
globular clusters), similarly high pressures must have been common
during the peak of globular cluster formation a few billion years
after the Big Bang.

\acknowledgements We thank the anonymous referee for their many useful
comments.  This paper makes use of the following ALMA data :
ADS/JAO.ALMA\#2011.0.00876.S. ALMA is a partnership of ESO
(representing its member states), NSF (USA) and NINS (Japan), together
with NRC (Canada) and NSC and ASIAA (Taiwan), in cooperation with the
Republic of Chile. The Joint ALMA Observatory is operated by ESO,
AUI/NRAO and NAOJ.  The National Radio Astronomy Observatory is a
facility of the National Science Foundation operated under cooperative
agreement by Associated Universities, Inc. K.E.J. acknowledges support
provided by the David and Lucile Packard Foundation through a Packard
Fellowship.



\end{document}